\documentclass[aps,pra,twocolumn,superscriptaddress,showpacs]{revtex4}
\usepackage{mathrsfs}
\usepackage{graphicx}
\usepackage{amsfonts}
\usepackage{amsmath}
\usepackage{amssymb}

\begin{document}

\title{Quantum nonlocality of massive qubits in a moving frame}

\author{Hong-Yi~Su}
 \affiliation{Theoretical Physics Division, Chern Institute of Mathematics, Nankai University,
 Tianjin 300071, People's Republic of China}

\author{Yu-Chun Wu}
 \affiliation{Key Laboratory of Quantum Information, University of Science and Technology of China, 230026 Hefei, People's Republic of China}

 \author{Jing-Ling~Chen}
 \email{chenjl@nankai.edu.cn}
 \affiliation{Theoretical Physics Division, Chern Institute of Mathematics, Nankai University,
 Tianjin 300071, People's Republic of China}
 \affiliation{Centre for Quantum Technologies, National University of Singapore,
 3 Science Drive 2, Singapore 117543}

 \author{Chunfeng~Wu}
 \affiliation{Pillar
of Engineering Product Development, Singapore University of
Technology and Design, 20 Dover Drive, Singapore 138682 }

\author{L.~C.~Kwek}
\email{kwekleongchuan@nus.edu.sg}
 \affiliation{Centre for Quantum Technologies, National University of Singapore,
 3 Science Drive 2, Singapore 117543}
 \affiliation{National Institute of Education and Institute of Advanced Studies,
 Nanyang Technological University, 1 Nanyang Walk, Singapore 637616}
 \affiliation{Institute of Advanced Studies, Nanyang Technological University, 60 Nanyang View
Singapore 639673}

\date{\today}

\begin{abstract}
We perform numerical tests on quantum nonlocality of two-level
quantum systems (qubits) observed by a uniformly moving
observer.  Under a suitable momentum setting, the quantum nonlocality of
two-qubit nonmaximally entangled states could be weakened drastically
by the Lorentz transformation allowing for the existence of local-hidden-variable
models, whereas three-qubit genuinely entangled states are
robust.  In particular, the generalized GHZ state remains nonlocal under arbitrary
Wigner rotation and the generalized W state could admit
local-hidden-variable models within a rather narrow range of parameters.
\end{abstract}

\pacs{03.65.Ud, 03.30.+p, 03.67.-a}

\maketitle

\section{Introduction}

Most issues in quantum information science focussed on
problems that elucidate and exemplify the difference between classical and
quantum mechanics within non-relativistic context.  However, in recent years, it has
been realized that fundamental notions in quantum information theory undergo
substantial revision under relativistic settings.
The theory of relativity requires a
physical quantity to be Lorentz-invariant.  Thus, there were several attempts~\cite{attemps}
to develop notions in quantum
information like measures for quantum entanglement and quantum teleportation fidelity
and so forth within
relativistic settings so that they remain Lorentz
invariant.  There is generally no consensus on such modifications needed in this respect~\cite{Peres}.
Clearly, the search for Lorentz-invariant properties in quantum systems is quite a
non-trivial task.

Bell's inequality~\cite{Bell,CHSH} can be regarded as a hybrid of
relativity and quantum mechanics. On one hand, its derivation adopts
the assumption of locality, an important feature in relativity. On
the other hand, an observable under measurements is described by the
Hermitian operator based on the standard techniques in quantum
mechanics to account for a measurable physical quantity. One may
observe that Lorentz-invariance, an important feature in relativity,
is explicitly missing in Bell's inequality.  This naturally leads to
the question on how quantum nonlocality adapts itself when Lorentz
invariance is taken into account. This is principal motivation
behind our current work.

To this end, we consider the Wigner rotation~\cite{Wigner}, a
significant relativistic effect related to the Lorentz
transformation. Before proceeding further, there is an important
consideration: viz. the relativistic counterpart of the spin
operator in quantum
mechanics~\cite{r-spin-operator,r-spin-operator-eg1,r-spin-operator-eg2}.
However, as we show in Sec.~III, insofar as quantum nonlocality is
concerned, we may somehow neglect this consideration for all
practical purposes for qubit systems.

It is convenient to recast the eigenstate of quantum systems as a
product of ``momentum state" and ``spin state":
$|\Psi\rangle=|\psi_{\rm mom}\rangle\otimes|\phi_{\rm spin}\rangle$.
Under relativistic settings, there is mixing between the spin and
momentum parts. Indeed, the relativistic quantum nonlocality of two-
and three-qubit maximally entangled spin states have been explored
in Refs.~\cite{QN-Max}. In these references, the momentum state is
essentially a product state and the Wigner rotation is equivalent to
a local unitary transformation of the spin state. Under such
condition, quantum nonlocality is anticipated to be
Lorentz-invariant as well as frame-independent. Authors in
Ref.~\cite{partial-entropy} also considered the two-qubit case where
the momentum state is entangled and showed that the partial entropy
drastically decreases with respect to large Wigner angles. In
Ref.~\cite{Huber}, the authors considered an alternative three-qubit
momentum setting with two genuinely entangled spin states (GHZ and W
states)  using a version of  multi-partite concurrence and derived
general conditions which have to be met for any classification of
multi-partite entanglement to be Lorentz-invariant.

In this work, we consider more general situations using Bell's
inequality. We first consider the two-qubit case in which the spin
part is a generalized GHZ state. We then investigate two
inequivalent classes of genuine three-qubit entangled states:
generalized GHZ and W states. Different momentum settings are also
compared and discussed.

The paper is organized as follows. In Sec.~II, we provide a
definition of the Wigner rotation and discuss its effect on multi-partite
quantum states. In Sec.~III, we discuss the possible candidates for
relativistic spin operators and their relationship to the usual
Pauli operator. In Sec.~IV, we present our main results on
relativistic quantum nonlocality of two and three qubits. We end the
paper with a summary.

\section{The Wigner Rotation}

The Wigner rotation can be understood algebraically as the consequence of
the non-associativity of the relativistic addition of velocities
(Einstein's addition). In group theory, the Wigner rotation is
rigorously defined as the little group~\cite{Wigner}:
\begin{eqnarray}
W(\Lambda,p)=L^{-1}(\Lambda p)\Lambda L(p),\label{wigner-rot}
\end{eqnarray}
where $L(p)$ denotes a standard Lorentz transformation that
transforms the particle from rest to four-momentum $p^\mu$, namely,
\begin{eqnarray}
p^\mu=L^\mu_\nu(p)k^\nu,\;\;k^\nu=(m,0,0,0).
\end{eqnarray}
[In this section, we use notations similar to those in
Ref.~\cite{Weinberg} and adopt natural units making $c=1$.]

The one-particle state with momentum $\vec{p}$ and spin $\sigma$ can
be expressed by
\begin{eqnarray}
|\vec{p},\sigma\rangle\equiv a^\dagger(\vec{p},\sigma)|{\rm
vac}\rangle,\label{one-particle-state}
\end{eqnarray}
with $|{\rm vac}\rangle$ the Lorentz-invariant vacuum state and
$a^\dagger(\vec{p},\sigma)$ the creation operator transforming under
the unitary $U(\Lambda)$ with the following rule:
\begin{eqnarray}
U(\Lambda)a^\dagger(\vec{p},\sigma)U^{-1}(\Lambda)=\sum_{\sigma'}D^j_{\sigma'\sigma}(W(\Lambda,p))a^\dagger(\vec{p}_\Lambda,\sigma'),\label{adagger-trans}
\end{eqnarray}
where $D^j_{\sigma'\sigma}(W(\Lambda,p))$ are elements of the
$(2j+1)$-dimensional representation of the Wigner rotation
$W(\Lambda,p)$, and $\vec{p}_\Lambda$ is the spatial component of
the transformed four-momentum $\Lambda p$. Thus, when the observer is
moving at a certain constant velocity $\vec{v}$, the one-particle
state (\ref{one-particle-state}) is transformed further to
\begin{eqnarray}
U(\Lambda)|\vec{p},\sigma\rangle=\sum_{\sigma'}D^j_{\sigma'\sigma}(W(\Lambda,p))|\vec{p}_\Lambda,\sigma'\rangle.\label{Wig-r}
\end{eqnarray}

We can interpret (\ref{Wig-r}) in the following manner: The observer
is at rest and the frame which contains the particle in the state
$|\vec{p},\sigma\rangle$ is moving. Then the Wigner angle arises
after three steps: (i) a particle with spin $\sigma$ is created at
rest from vacuum, we get the state $|0,\sigma\rangle$; (ii) the
frame which contains the particle moves in the velocity
$\vec{u}=\vec{p}/p^0$, thus the one-particle state
$|0,\sigma\rangle$ is transformed to $|\vec{p},\sigma\rangle$ by
$U(L(p))$; (iii) the frame further moves in the velocity $\vec{v}$,
i.e. the second Lorentz transformation $U(\Lambda)$ acts on state
$|\vec{p},\sigma\rangle$. According to (\ref{wigner-rot}), two
successive Lorentz transformations $L(p)$ and $\Lambda$ are equal to
a single $L(\Lambda p)$ combined with the Wigner rotation $W$.
Therefore, the scenario of a moving particle observed in the moving
frame is elegantly equivalent to that of an observer at rest who
observes a successively Lorentz-transformed particle.

The multipartite state is expressed by
\begin{eqnarray}
|\vec{p}_1,\sigma_1;\vec{p}_2,\sigma_2\cdots\rangle\equiv
a^\dagger(\vec{p}_1,\sigma_1)a^\dagger(\vec{p}_2,\sigma_2)\cdots|{\rm
vac}\rangle,\label{multi-particle-state}
\end{eqnarray}
As observed in a moving frame, each $a^\dagger(\vec{p}_k,\sigma_k)$
transforms according to (\ref{adagger-trans}). The
Lorentz-transformed state is found to be
\begin{eqnarray}
&&U(\Lambda)|\vec{p}_1,\sigma_1;\vec{p}_2,\sigma_2\cdots\rangle\nonumber\\
&&\;\;\;\;=\sum_{\sigma_1'\sigma_2'\cdots}D^{j_1}_{\sigma_1'\sigma_1}(W(\Lambda,p_1))
D^{j_2}_{\sigma_2'\sigma_2}(W(\Lambda,p_2))\cdots\nonumber\\
&&\;\;\;\;\;\;\;\;\;\;\;\;\;\;\;\;\;\;\;\;\;\;\;\;\;\;\;\;\;\;\;\;\;\;\;\;\;\;\;\;
\times|\vec{p}_{1\Lambda},\sigma_1';\vec{p}_{2\Lambda},\sigma_2'\cdots\rangle.\label{multi-particle-state-trans}
\end{eqnarray}
For the sake of convenience later, we can also separate momentum
$\vec{p}_k$ explicitly from spin $\sigma_k$ so that the multipartite
state is effectively a product of momentum and spin. For instance,
Eq.~(\ref{multi-particle-state-trans}) becomes
\begin{eqnarray}
&&U(\Lambda)|\vec{p}_1,\vec{p}_2\cdots\rangle\otimes|\sigma_1,\sigma_2\cdots\rangle\nonumber\\
&&\;\;\;\;=|\vec{p}_{1\Lambda},\vec{p}_{2\Lambda}\cdots\rangle\nonumber\\
&&\;\;\;\;\;\;\;\;\otimes\sum_{\sigma_1'\sigma_2'\cdots}D^{j_1}_{\sigma_1'\sigma_1}(W(\Lambda,p_1))
D^{j_2}_{\sigma_2'\sigma_2}(W(\Lambda,p_2))\cdots\nonumber\\
&&\;\;\;\;\;\;\;\;\;\;\;\;\;\;\;\;\;\;\;\;\;\;\;\;\;\;\;\;\;\;\;\;\;\;\;\;\;\;\;\;\;\;\;\;\;\;\;\;\;\;\;\;\;\;\;
\times|\sigma_1',\sigma_2'\cdots\rangle.
\end{eqnarray}
Note that if we focus on the spin state
$|\sigma_1,\sigma_2\cdots\rangle$, the Wigner rotation can be
regarded as a local unitary transformation since the little group
$W(\Lambda,p_k)$ and its the representation
$D^{j_k}(W(\Lambda,p_k))$ are unitary.

The little group $W(\Lambda,p)$ for massive particles is $SO(3)$.
Representations of this little group have been systematically
studied using the method of induced
representations~\cite{induced-rep}. In the following context, we
restrict ourselves to two-level particles (qubits). The
two-dimensional representations $D^{1/2}(W(\Lambda,p_k))$ are
employed and the generators are Pauli matrices $\vec{s}$. For our
purpose, the particles are moving with velocity $\vec{u}_k$ in the
$yz$-plane and the observer is moving with velocity $\vec{v}$ along
the $x$-axis, then the two-dimensional representation is found to
be~\cite{Weinberg}
\begin{eqnarray}
D^{j_k=\frac{1}{2}}(W(\Lambda,p_k))
=\cos\frac{\Omega(\vec{u}_k,\vec{v})}{2}+i\vec{s}\cdot\vec{n}_k\sin\frac{\Omega(\vec{u}_k,\vec{v})}{2},
\end{eqnarray}
with $\vec{n}_k=\vec{v}\times\vec{u}_k/(|\vec{v}|\cdot|\vec{u}_k|)$.
The Wigner angle $\Omega(\vec{u}_k,\vec{v})$ is calculated by
\begin{eqnarray}
\Omega(\vec{u}_k,\vec{v})=\arctan\frac{\sinh\xi\sinh\zeta}{\cosh\xi+\cosh\zeta},
\end{eqnarray}
with
\begin{eqnarray}
\cosh\xi=\frac{1}{\sqrt{1-|\vec{u}_k|^2}},\;\;
\cosh\zeta=\frac{1}{\sqrt{1-|\vec{v}|^2}}.
\end{eqnarray}
Note that the Wigner angle is zero as $|\vec{u}_k|=0$ or
$|\vec{v}|=0$, and goes up to $\pi/2$ as $|\vec{u}_k|$ and
$|\vec{v}|$ approach the speed of light $c=1$.

Suppose the initial state can be rewritten as a product of momentum
and spin states, i.e.,
\begin{eqnarray}
|\Psi\rangle=|\psi_{\rm mom}\rangle\otimes|\phi_{\rm spin}\rangle,
\end{eqnarray}
or
\begin{eqnarray}
\rho=\rho_{\rm mom}\otimes\rho_{\rm spin},
\end{eqnarray}
with
\begin{eqnarray}
\rho_{\rm mom}=|\psi_{\rm mom}\rangle \langle\psi_{\rm mom}|,\;\;
\rho_{\rm spin}=|\psi_{\rm spin}\rangle \langle\psi_{\rm spin}|.
\end{eqnarray}
Here $\rho_{\rm mom}$ and $\rho_{\rm spin}$ could be entangled states.
To see the dependence of  the Wigner rotation on momentum, the state
observed by the moving observer becomes
\begin{eqnarray}
\rho'=U(\Lambda)\rho U(\Lambda)^{-1}.
\end{eqnarray}
This transformed state may not generally be in the form of product
$\rho'_{\rm mom}\otimes\rho'_{\rm spin}$ and the reduced spin state
\begin{eqnarray}
\rho'_{\rm spin}={\rm tr}_{\rm mom} \rho'
\end{eqnarray}
is a mixed state. Since it has been proved that there exist some
mixed states that bear local-hidden-variable models~\cite{Werner},
the Wigner rotation of quantum states is non-trivial in the study of
quantum nonlocality.

\section{The Relativistic Observables}

There have been several proposals for the relativistic spin
operator~\cite{r-spin-operator-eg1,r-spin-operator-eg2}, derived
under very  different physical requirements and not mutually
equivalent to one another. For instance, deriving from relativistic
center of mass, one obtains a typical relativistic spin
operator~\cite{r-spin-operator-eg1}:
\begin{eqnarray}
\hat{a}=\frac{(\sqrt{1-|\vec{w}_k|^2}\vec{a}^\perp+\vec{a}^\parallel
)\cdot\vec{s}}{\sqrt{1+(\vec{w}_k\cdot\vec{a})^2-|\vec{w}_k|^2}},\label{rel-spin}
\end{eqnarray}
where $\vec{a}=\vec{a}^\perp+\vec{a}^\parallel$ is the measuring
direction with components $\vec{a}^\perp$ and $\vec{a}^\parallel$
respectively perpendicular and parallel to the direction of the
composite velocity $\vec{w}_k$ of $\vec{u}_k$ and
$\vec{v}$~\cite{einstein-add}.

One can verify that eigenvalues of this relativistic spin operator
are $\pm1$.
In fact, by rotating along a certain direction, any normalized
relativistic spin operator which is in the form of linear
combination of Pauli matrices coincides with the nonrelativistic
spin operator $\vec{a}\cdot\vec{s}$. To see this clearly, let us
take another look at the operator (\ref{rel-spin}). As mentioned
above, this is a relativistic spin operator measured in the
direction $\vec{a}$. From the nonrelativistic viewpoint,
(\ref{rel-spin}) is effectively a nonrelativistic spin operator
$\vec{a'}\cdot\vec{s}$ measured in the direction $\vec{a'}$. In
principle, there must be a unitary $U$ that rotates $\vec{a'}$ to
$\vec{a}$, i.e.,
\begin{eqnarray}
\vec{a'}\cdot\vec{s}=U\vec{a}\cdot\vec{s}\;U^{-1}.
\end{eqnarray}
This is crucial in the relativistic quantum nonlocality. This fact
implies that instead of the relativistic ones we can adopt the
nonrelativistic spin operators to obtain the quantum violation of
Bell's inequality. In other words, if one uses the relativistic spin
operators measured along a certain set of directions to calculate
the quantum value, then one can obtain the same value by using the
nonrelativistic ones, with measuring directions modified by some
unitary $U$.

The $M$-setting $N$-qubit Bell inequality $I_N$ can be written in
the correlational form:
\begin{eqnarray}
I_N=\sum_{i_1,i_2,...,i_N=0}^M T_{i_1i_2...i_N} Q_{i_1i_2...i_N}\leq
1,
\end{eqnarray}
where $T_{i_1i_2...i_N}$ are coefficients and $Q_{i_1i_2...i_N}$ are
correlation functions defined by
\begin{eqnarray}
Q_{i_1i_2...i_N}={\rm tr}[\rho_{\rm spin}
\vec{a}_{i_1}\cdot\vec{s}\otimes\vec{a}_{i_2}\cdot\vec{s}\otimes\cdots\vec{a}_{i_N}\cdot\vec{s}],
\end{eqnarray}
with $\vec{a}_{i_k}$ the $i_k$-th measuring direction of the $k$-th
qubit, if $i_k\neq0$. When $i_k=0$, this means no measurement is
performed on the $k$-th qubit, thus the correlation function is
modified by substituting the identity $\openone$ for spin operator
$\vec{a}_{i_k}\cdot\vec{s}$.

Following the analysis above, the quantum value $I_N$ with respect
to some settings $\{\vec{a}_{i_1},\vec{a}_{i_2},...,\vec{a}_{i_N}\}$
using relativistic spin operators is the same to that with respect
to modified settings
$\{\vec{a'}_{i_1},\vec{a'}_{i_2},...,\vec{a'}_{i_N}\}$ using
nonrelativistic ones. Therefore, as far as qubit systems are
concerned, the usual Pauli operator $\vec{s}$ is adequate to study
the quantum nonlocality.

\emph{Remark.---} An alternative way to describe a relativistic spin
measurement is to utilize the helicity operator acting on the total
state~\cite{Czachor1}. However, for massive particles, there is a
subtlety on this choice of helicity observables, as raised by
Czachor in Ref.~\cite{Czachor2} in which the Pauli-Lubanski operator
is projected on a null direction instead of the time direction,
leaving the entanglement of particles unchanged under a Lorentz
boost. In the present work, we follow the line of Ref.~\cite{Peres}:
one has the reduced spin state defined as a trace-out of momenta,
before measuring the particle spin. Accordingly, detectors in an
experiment must distinguish only polarizations, and the information
on motion is discarded.

\section{Relativistic Quantum Nonlocality of Two and Three Qubits}

Recalling the analysis in Sec.~II, the observed spin state for the
moving observer is a Lorentz-transformed one:
\begin{eqnarray}
\rho_{\rm spin}\rightarrow \rho'_{\rm spin}.
\end{eqnarray}
To test this state, we can readily use the Bell inequalities which
have been employed to study nonlocality in nonrelativistic quantum
mechanics, since the derivation of Bell inequalities is inherently
in the relativistic treatment.

For two qubits, we consider the CHSH inequality~\cite{CHSH}
\begin{eqnarray}
I_2=\frac{1}{2}\biggr( Q_{11}+Q_{12}+Q_{21}-Q_{22} \biggr)\leq 1,
\end{eqnarray}
with $Q_{ij}={\rm tr}[\rho
\vec{a}_i\cdot\vec{s}\otimes\vec{a}_j\cdot\vec{s}]$. The initial
state we consider is
\begin{eqnarray}
&&|\Psi_1\rangle=\biggr( \cos\theta_m|
\vec{p}_1,\vec{p}_2\rangle+\sin\theta_m|-\vec{p}_1,-\vec{p}_2\rangle \biggr)\nonumber\\
&&\;\;\;\;\;\;\;\;\;\;\;\;\;\;\;\;\;\;\;\;\;\;\;\;\;\;\;\;\otimes
\biggr( \cos\theta_s|00\rangle+\sin\theta_s|11\rangle
\biggr),\label{twoqubit}
\end{eqnarray}
with two different momentum settings
\begin{eqnarray}
{\rm p_1}=-{\rm p_2}=(0,0,1),\label{twoqubit-setting1}
\end{eqnarray}
and
\begin{eqnarray}
{\rm p_1}={\rm p_2}=(0,0,1).\label{twoqubit-setting2}
\end{eqnarray}
Here ${\rm p}_k=\vec{p}_k/|\vec{p}_k|$ indicates the moving
direction of the $k$-th qubit (i.e., ${\rm p}_k$ is proportional to
$\vec{u}_k$ in Sec.~II). In the former setting, two qubits are in
the superposition of moving in opposite directions, while in the
latter they are in the superposition of moving in the same direction
($z$-axis or the opposite). For simplicity, we take
$|\vec{p}_1|=|\vec{p}_2|$.

\begin{figure}[tbp]
\includegraphics[width=80mm]{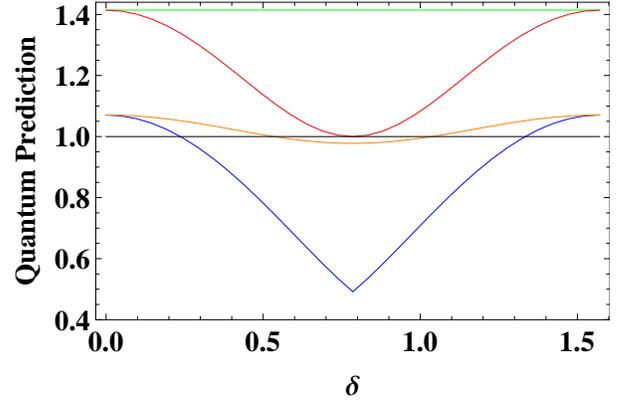}\\
\caption{(Color online) The variation of quantum values $I_2$ with
respect to the Wigner rotations of the initial state
(\ref{twoqubit}) in different momentum settings
(\ref{twoqubit-setting1}) and (\ref{twoqubit-setting2}). The red and
blue curves correspond to parameters $\{\theta_m,\theta_s\}$ taking
$\{\frac{\pi}{4},\frac{\pi}{4}\}$ and
$\{\frac{\pi}{4},\frac{\pi}{16}\}$ in (\ref{twoqubit-setting1}),
respectively; and the green and orange curves correspond to
parameters $\{\theta_m,\theta_s\}$ taking
$\{\frac{\pi}{4},\frac{\pi}{4}\}$ and
$\{\frac{\pi}{4},\frac{\pi}{16}\}$ in (\ref{twoqubit-setting2}),
respectively.}\label{fig1}
\end{figure}

As the experimentalist moves in the opposite $x$-axis (or
equivalently, the experimentalist stays in his rest frame and the
qubits system moves in the $x$-axis), the observed state is a
transformed one by the Wigner rotation of (\ref{twoqubit}).
Fig.~\ref{fig1} shows the quantum values of $I_2$ with respect to
various Wigner angles. For the momentum setting
(\ref{twoqubit-setting1}) (see the red and blue curves), there would
be a region where local-hidden-variable models are admitted, unless
either the entanglement degree of the momentum part is small or the
spin state is maximally entangled. For the momentum setting
(\ref{twoqubit-setting2}) (see the green and orange curves), we have
similar results, except that (i) the effect of the Wigner rotation
is weaker than that in (\ref{twoqubit-setting1}), and (ii) the
Wigner rotation does not affect the quantum values when the spin
state is maximally entangled.

For three qubits, we use the following inequality proposed in
Ref.~\cite{Chen}:
\begin{eqnarray}
I_3=\frac{1}{3}\biggr( -Q_{111}+Q_{221}+Q_{212}+Q_{122}-Q_{222}
-Q_{110}\nonumber\\
-Q_{120}-Q_{210}
-Q_{101}-Q_{102}-Q_{201}-Q_{011}\nonumber\\
-Q_{012}-Q_{021}+Q_{200}+Q_{020}+Q_{002} \biggr)\leq 1,\nonumber\\
\end{eqnarray}
with $Q_{ijk}={\rm tr}[\rho
\vec{a}_i\cdot\vec{s}\otimes\vec{a}_j\cdot\vec{s}\otimes\vec{a}_k\cdot\vec{s}]$,
and subscript ``0" indicating that no measurement is performed on
the corresponding qubit. For instance, $Q_{ij0}={\rm tr}[\rho
\vec{a}_i\cdot\vec{s}\otimes\vec{a}_j\cdot\vec{s}\otimes\openone]$.
The initial states we consider are
\begin{eqnarray}
&&|\Psi_2\rangle=\biggr( \cos\theta_m|\vec{p}_1, \vec{p}_2,
\vec{p}_3\rangle+\sin\theta_m|-\vec{p}_1, -\vec{p}_2,
-\vec{p}_3\rangle
\biggr)\nonumber\\
&&\;\;\;\;\;\;\;\;\;\;\;\;\;\;\;\;\;\;\;\;\;\;\;\;\;\;\;\otimes
\biggr(
\cos\theta_s|000\rangle+\sin\theta_s|111\rangle \biggr),\label{threequbit-GHZ}\\
&&|\Psi_3\rangle=\biggr( \cos\theta_m|\vec{p}_1, \vec{p}_2,
\vec{p}_3\rangle+\sin\theta_m|-\vec{p}_1, -\vec{p}_2,
-\vec{p}_3\rangle
\biggr)\nonumber\\
&&\;\;\;\;\;\;\;\;\;\;\;\otimes \biggr(
\sin\theta_s\cos\phi_s|001\rangle+\sin\theta_s\sin\phi_s|010\rangle\nonumber\\
&&\;\;\;\;\;\;\;\;\;\;\;\;\;\;\;\;\;\;\;\;\;\;\;\;\;\;\;\;\;\;\;\;\;\;\;\;\;\;\;\;\;\;\;\;\;\;\;\;\;\;
+\cos\theta_s|100\rangle \biggr),\label{threequbit-W}
\end{eqnarray}
with
\begin{eqnarray}
&&{\rm p}_1=(0,0,1),\nonumber\\
&&{\rm p}_2=(0,{\sqrt{3}}/{2},-{1}/{2}),\nonumber\\
&&{\rm p}_3=(0,-{\sqrt{3}}/{2},-{1}/{2}),\label{threequbit-setting1}
\end{eqnarray}
with ${\rm p}_k=\vec{p}_k/|\vec{p}_k|$ and
$|\vec{p}_1|=|\vec{p}_2|=|\vec{p}_3|$, similar to the two-qubit
case. Note that the spin states of (\ref{threequbit-GHZ}) and
(\ref{threequbit-W}) belong to two inequivalent classes of genuine
three-qubit entangled states: generalized GHZ and W states.

\begin{figure}[tbp]
\includegraphics[width=80mm]{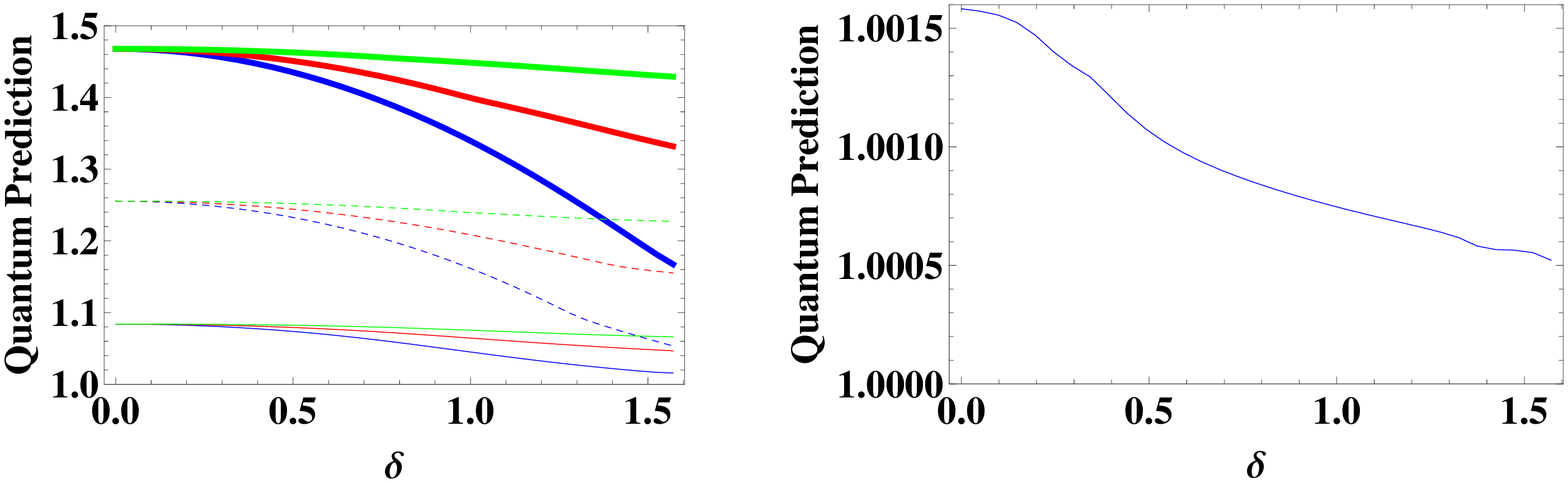}\\
\caption{(Color online) The variation of quantum values $I_3$ with
respect to the Wigner rotations of the initial state
(\ref{threequbit-GHZ}) in the momentum setting
(\ref{threequbit-setting1}). Left: The colors $\{{\rm
blue,red,green}\}$ correspond to parameter
$\theta_m=\{\frac{\pi}{4},\frac{\pi}{8},\frac{\pi}{16}\}$,
respectively; the line styles $\{{\rm thick,dashed,thin}\}$
correspond to parameter
$\theta_s=\{\frac{\pi}{4},\frac{\pi}{8},\frac{\pi}{16}\}$,
respectively. Right: The curve corresponds to
$\{\theta_m,\theta_s\}=\{\frac{\pi}{4},\frac{\pi}{128}\}$.}\label{fig2}
\end{figure}

\begin{figure}[tbp]
\includegraphics[width=80mm]{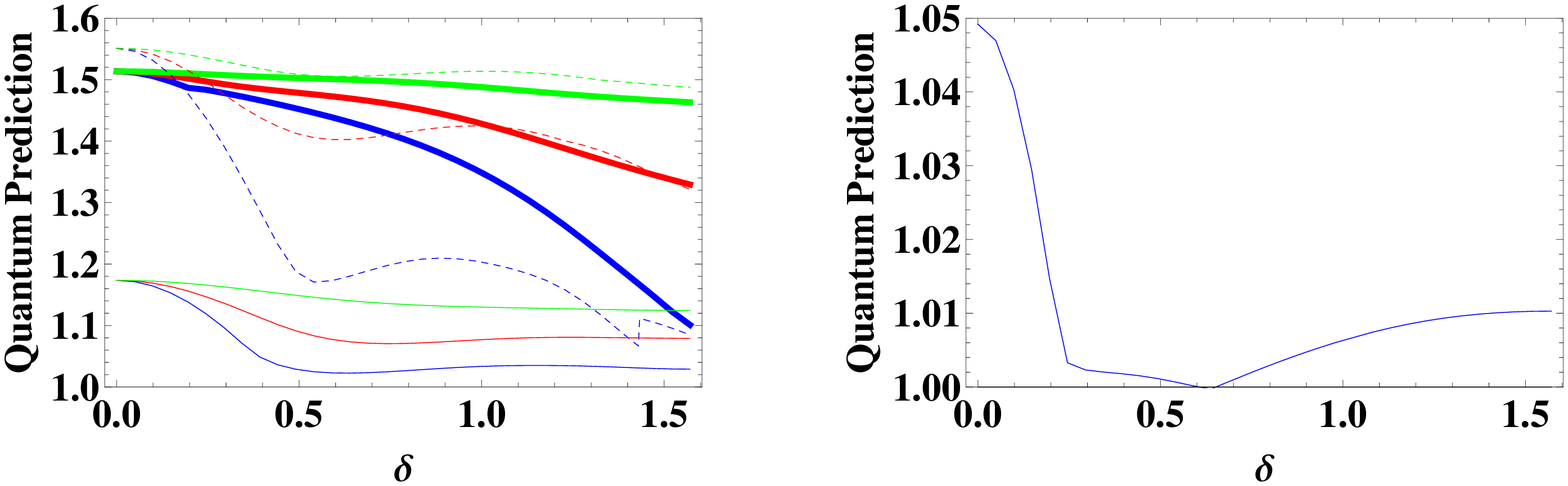}\\
\caption{(Color online) The variation of quantum values $I_3$ with
respect to the Wigner rotations of the initial state
(\ref{threequbit-W}) in the momentum setting
(\ref{threequbit-setting1}). Left: The colors $\{{\rm
blue,red,green}\}$ correspond to parameter
$\theta_m=\{\frac{\pi}{4},\frac{\pi}{8},\frac{\pi}{16}\}$,
respectively; the line styles $\{{\rm thick,dashed,thin}\}$
correspond to parameter pair
$\{\theta_s,\phi_s\}=\biggr\{\{\arccos\frac{1}{\sqrt{3}},\frac{\pi}{4}\},\{\frac{7\pi}{16},\frac{\pi}{4}\},\{\frac{7\pi}{16},\frac{\pi}{16}\}\biggr\}$,
respectively. Right: The curve corresponds to
$\{\theta_m,\theta_s,\phi_s\}=\{\frac{\pi}{4},\frac{15\pi}{32},\frac{\pi}{32}\}$.}\label{fig3}
\end{figure}

Fig.~\ref{fig2} (left) and Fig.~\ref{fig3} (left) show the quantum
values of $I_3$ with respect to various Wigner angles for several
typical parameters taken in the initial states
(\ref{threequbit-GHZ}) and (\ref{threequbit-W}), respectively. It is
found that for given parameters in the spin states, the Wigner
rotation weakens quantum nonlocality the most as the momentum states
are maximally entangled. Thus in Fig.~\ref{fig2} (right) we further
take $\theta_m=\frac{\pi}{4}$ (maximally entangled momentum state)
and $\theta_s=\frac{\pi}{128}$ (close to the separable spin state
$|000\rangle$), and then find that quantum values are still larger
than 1.

In Fig.~\ref{fig3} (left) we consider three types of generalized W
states: (i) all three components
$\{|001\rangle,|010\rangle,|100\rangle\}$ have equal weights (see
three thick curves), (ii) one component is relatively smaller than
the others (see three dashed curves), and (iii) one component is
relatively larger than the others (see three thin curves). Among
them, type (ii) is close to bi-separable state
$|0\rangle\otimes(|01\rangle+|10\rangle)/\sqrt{2}$, type (iii) is
close to tri-separable state $|001\rangle$. In Fig.~\ref{fig3}
(right) we further take $\theta_m=\frac{\pi}{4}$,
$\theta_s=\frac{15\pi}{32}$ and $\phi_s=\frac{\pi}{32}$, and then
find that quantum values are larger than 1 for almost the whole
region. The minimum is approximately 0.9997 near the point
$\delta\approx 0.64$.

Moreover, the GHZ and W states as the spin states in
(\ref{threequbit-GHZ}) and (\ref{threequbit-W}) are the only two
classes of genuine three-qubit entangled states~\cite{ent-classes};
the other genuine entangled states are local unitary (LU) equivalent
to one of them. Therefore, it is reasonable to draw a conclusion
that quantum nonlocality of genuine three-qubit entangled states is
robust against the Lorentz transformation.

\begin{figure}[tbp]
\includegraphics[width=80mm]{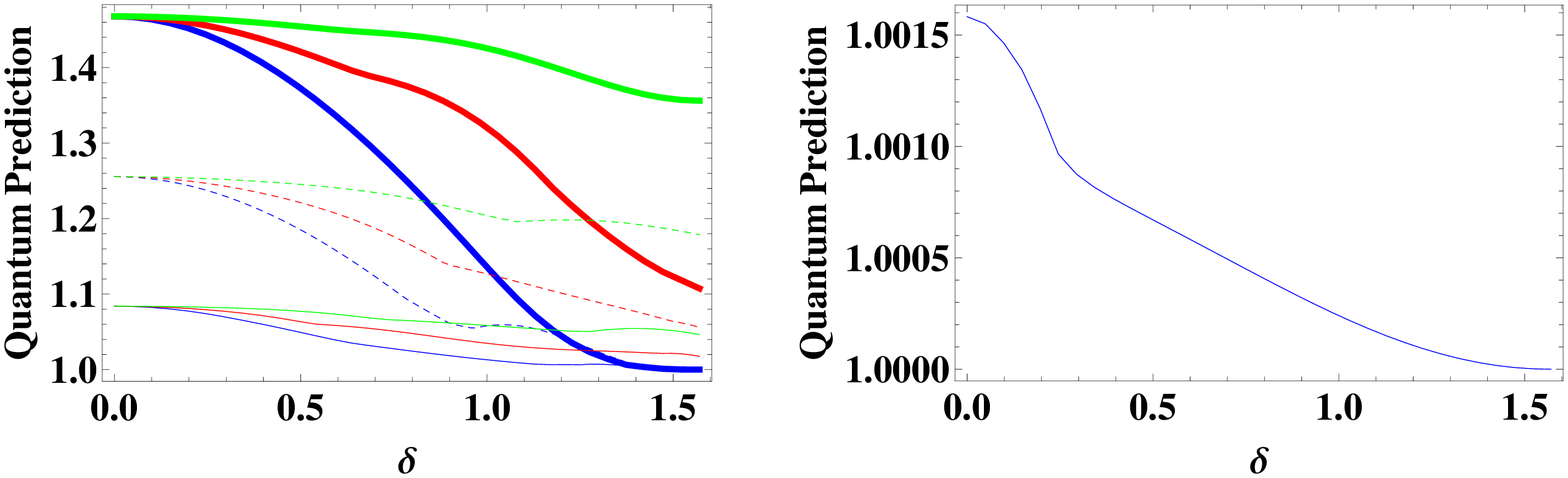}\\
\caption{(Color online) The variation of quantum values $I_3$ with
respect to the Wigner rotations of the initial state
(\ref{threequbit-GHZ}) in the momentum setting
(\ref{threequbit-setting2}). Left: The colors $\{{\rm
blue,red,green}\}$ correspond to parameter
$\theta_m=\{\frac{\pi}{4},\frac{\pi}{8},\frac{\pi}{16}\}$,
respectively; the line styles $\{{\rm thick,dashed,thin}\}$
correspond to parameter
$\theta_s=\{\frac{\pi}{4},\frac{\pi}{8},\frac{\pi}{16}\}$,
respectively. Right: The curve corresponds to
$\{\theta_m,\theta_s\}=\{\frac{\pi}{4},\frac{\pi}{128}\}$.}\label{fig4}
\end{figure}

We must stress that this conclusion is drawn by a proper selection
of momentum magnitude and directions. Under the Lorentz
transformation, the inevitable coupling of momentum and spin results
in the relativistic quantum nonlocality of spin states also
sensitively depending on details of momentum. To see this, let us
change momentum directions in (\ref{threequbit-GHZ}) to
\begin{eqnarray}
{\rm p}_1={\rm p}_2={\rm p}_3=(0,0,1).\label{threequbit-setting2}
\end{eqnarray}
The corresponding quantum values are shown in Fig.~\ref{fig4}. It is
obvious that the curves are different from those in Fig.~2.

However, it is also interesting that in this momentum setting the
Lorentz-transformed state still remains nonlocal under the Lorentz
transformation. Whether other types of entangled momentum and spin
parts (for instance, partially entangled states) remain nonlocal
under the Lorentz transformation, and if not, how the nonlocality is
weakened with respect to various parameters in transformations are
also intriguing questions subsequently.

\section{Summary}

To investigate relativistic quantum nonlocality, we have taken into
account the composite motion of both spins and the observer. This
motion is non-trivial and will cause the Wigner rotation of particle
states. We have shown that quantum nonlocality of two-qubit states
could be drastically weakened if the entanglement degree is not
maximal. In the three-qubit case, however, we have shown that
quantum nonlocality of genuinely entangled states remains nonlocal
with respect to almost arbitrary Wigner angles. Moreover, we have
also pointed out that one should carefully consider the details of
particle momentum, since spin is inevitably coupled to momentum
under the Lorentz transformation.

Here are a few words before ending the paper. Physically, the
momentum setting in (\ref{twoqubit-setting1}) or
(\ref{threequbit-setting1}) describes a particle that decays into
several subparticles traveling uniformly in space with two
possibilities. An alternative momentum setting as taken in
(\ref{twoqubit-setting2}) or (\ref{threequbit-setting2}) describes a
bunch of beam in which particles travel in the same direction,
positive or negative $z$-axis. These two settings may be more
feasible than the others in the experimental state preparation in
testing relativistic quantum nonlocality.

\begin{acknowledgments}
J.L.C. is supported by National Basic Research Program (973 Program)
of China under Grant No.\ 2012CB921900 and the NSF of China (Grant
Nos.\ 10975075 and 11175089). Y.C.W. acknowledges the support of the
National Basic Research Program of China (Grants No. 2011CBA00200
and No. 2011CB921200)and the National Natural Science Foundation of
China (Grant No. 11275182, 60921091). This work is also partly
supported by the National Research Foundation and the Ministry of
Education of Singapore.
\end{acknowledgments}

\end{document}